\documentstyle[12pt]{article}
\def\be{\begin{equation}}
\def\ee{\end{equation}}

\begin{document}
\title{On the Holographic $S$--matrix}

\author{
I.Ya.Aref'eva\\
{\it Steklov Mathematical Institute, Russian Academy of Sciences}\\
{\it Gubkin St.8, GSP-1, 117966, Moscow, Russia}\\
{\it Centro Vito Volterra, Universita di Roma Tor Vergata, Italy}\\
arefeva@genesis.mi.ras.ru}
\date {$~$}
\maketitle
\begin {abstract}
The recent proposal by Polchinski and Susskind for the holographic flat
space $S$--matrix is discussed.
By using  Feynman diagrams  we argue that in principle all
the information about the $S$--matrix in the interacting field theory in
the bulk of the anti-de Sitter space is encoded into the
data on the timelike boundary. The problem of locality of interpolating
field is discussed and it is suggested that the interpolating field lives
in a quantum Boltzmannian Hilbert space.
\end{abstract}

\newpage
According to the holographic principle \cite{tHo,Sus1} one should
describe a field theory on a manifold $M$ which includes gravity
 by a theory which lives on the
boundary of $M$.  Two prominent examples of the holography are the Matrix
theory \cite{BFSS} and the AdS/CFT correspondence \cite{Mal,GKP,Wit}.
The relation between quantum gravity in the anti-de Sitter space
and the gauge theory on the boundary could be useful for better
understanding of both
theories. In principle CFT
 might teach us about quantum gravity in the bulk of AdS.
Correlation functions in the Euclidean formulation
are the subject of intensive study (see for example \cite{AV}-\cite{IA}).
 The AdS/CFT correspondence
in the Lorentz formulation is considered in \cite{BKL}-\cite{BR}.

Recently Polchinski \cite{Pol} and Susskind \cite{Sus2} have proposed an
expression for the $S$--matrix in  flat spacetime in terms of
the large $N$ limit of the  gauge theory living on the boundary
of the AdS space. A related derivation is given in \cite{BGL}.
In this note we discuss this proposal.

It was suggested \cite{BKL,BDHM} that a quantum field $\Phi(t,{\bf x})$
in the bulk of $AdS_5$ and the corresponding
operator ${\cal O}(t,{\hat {\bf x}})$
in the gauge theory at the boundary are related by a simple formula
\be
\lim_{r\to\infty}r^4\Phi(t,{\bf x})={\cal O}(t,{\hat {\bf x}})
\label{AdS1}
\ee
Such a formula might be interpreted from two different points of view.
 One can assume that  the field ${\cal O}(t,{\hat {\bf x}})$
in CFT  is known and study what could be a field $\Phi(t,{\bf x})$
in the bulk. Or one can start from a field  in the bulk
of AdS and study its limiting behaviour at the boundary.
In this note we are interested in the discussion
of the latter approach  which perhaps can be helpful for a clarification
of the holographic principle.

We consider the following question.
Let us take a simple model of QFT in the flat 5--dimensional spacetime, say
the massless scalar field with $\Phi^3$ interaction. Can we get the
ordinary $S$--matrix for this model if we start from a quantum field theory in
the bulk of AdS and use the data on the time like boundary of AdS in the flat
space limit? We will see that an important point here is to answer
to the question what do we mean under the quantum field theory
in the bulk of AdS? In other words, how to quantize the  $\Phi^3$
theory in AdS if we want to reproduce the flat space $S$-matrix
by using only "holographical" data on the timelike infinity of AdS?

First we remind the standard formulae of the scattering theory in the
5--dimensional  Minkowski spacetime.
 There are four different formulations
of $S$-matrix on the flat spacetime,
see for example \cite{BogSh,LD,AFS}. They include the {\it in-out} formalism,
the LSZ formula, the Feynman diagrams and the on-shell functional
integral representation. It is well known that all of them are equivalent
in the flat spacetime. However in a curve background they could
lead to different prescriptions.

We consider a massless scalar field  with the action
\be
I=\int d^5x\left\{-{1\over2}\,(\partial_\mu\Phi)^2-V(\Phi)\right\}
\ee
where $V(\phi)=\lambda_1\phi^3+\lambda_2\phi^4+\dots$.
In the {\it in-out} formalism one defines an interpolating field
\begin{equation}
a({\bf k},x^0)=i\int d^4x(F_{{\bf k}}\partial_0\Phi
-\partial_0 F_{{\bf k}}\Phi)\label{H1}
\end{equation}
and in--out annihilation operators
\begin{equation}
a_{out,in}({\bf k})=\lim_{x^0\to\pm\infty}a({\bf k},x^0)
\label{H2}
\end{equation}
Here $F_{\bf k}$ is the solution of the wave equation with a positive
frequency,
$$\Box F_{{\bf k}}=0\ ,$$
$$F_{{\bf k}}(x)=e^{i\omega({\bf
k})x^0-i{\bf kx}}/(2\pi)^2\sqrt{2\omega({\bf k})}\ ,\quad
\omega({\bf k})=|{\bf k}|$$
$S$--matrix is
\begin{equation}
\langle
{\bf p}_1,\dots,{\bf p}_m; out |{\bf k}_1,
\dots,{\bf k}_n;in\rangle= \langle
a_{out}({\bf p}_1)\dots a_{out}({\bf p}_m)a^*_{in} ({\bf k}_1)\dots
a^*_{in}({\bf k}_n)\rangle
\label{H3}
\end{equation}

One can prove that (\ref{H3}) can be
written in the following standard form (the LSZ formula)
\begin{equation}
\langle a_{out}({\bf p}_1)\dots a^*_{in}({\bf k}_n)\rangle=
\int\prod^m_{i=1}F_{{\bf p}_i}(x_i)d^5x_i\prod^n_{j=1}F^*_{{\bf k}_j}(y_j)
d^5y_j
\Box_{x_1}\dots\Box_{y_n}G(x_1,\dots,y_n)+...
\label{H4}
\end{equation}
Here $G$ is the Feynman Green function,
\begin{equation}
G(x_1,\dots,x_n)=\langle T(\Phi(x_1)\dots\Phi(x_n))\rangle
\label{H5}
\end{equation}
This can be  can be represented in the perturbation theory as
\be
\langle T(\Phi(x_1)\dots\Phi(x_n))\rangle=
\langle T(\Phi_0(x_1)\dots\Phi_0(x_n)S)S^*\rangle
\label{H5a}
\ee
where

\begin{equation}
S=T\exp\left[-i\int d^5x ~V(\Phi_0)\right]\label{H6}
\end{equation}
and $\Phi_0(x)$ is the free field.
It is the form (\ref{H4})--(\ref{H6}) that is usually used for
computations  in perturbation theory and it can be expressed by using
the Feynman diagrams.

Now let us consider this quantum field theory in the $AdS_5$ spacetime.
The universal anti--de Sitter space is conformal to one half of the
Einstein static cylinder.
The metric is
\begin{equation}
ds^2=R^2\cosh^2\chi[-d\tau^2+d\sigma^2+\sin^2\sigma d\Omega]\label{H7}
\end{equation}
where
$\sigma=2\hbox{ arctan }e^\chi-{1\over2}\,\pi$
and the Penrose diagram is shown on Fig. 1
There are  null geodesics and ${\cal T}$ is time like infinity.

The action for the scalar field in $AdS_5$ is
\begin{equation}
I=\int d^5x\sqrt g\left\{-{1\over2}\,g^{\mu\nu}\partial_\mu\Phi
\partial_v\Phi-V(\Phi)\right\}\label{H7a}
\end{equation}
Now we have to quantize the field and try to repeat the main steps of the
scattering theory (\ref{H2})--(\ref{H5}). Moreover we want to study the
holographic principle and therefore we shall try to encode the data on
the scattering matrix into a theory on the timelike infinity ${\cal T}$.

One of problems we immediately get in the attempt to build a quantum
theory in $AdS$ space is that there exists no Cauchy surface in the
space.
It is clear from Fig. 1 that for any spacelike surface (for example a
horizontal section on Fig. 1) one can find null geodesics which never
intersect the surface.
So we have to restrict ourself to build a quantum theory only in a
region of $AdS$ space. This might be enough since our goal is to study
the flat space limit.

$AdS$ behaves like a cavity with reflecting walls \cite{Sus2}
and there are no
ordinary scattering states in $AdS$ space. It was proposed in
\cite{Pol,Sus2} to introduce sources and detectors on the boundary. Then in
the flat limit one can get the flat spacetime scattering amplitude if one
holds the external proper momenta fixed as $R\to\infty$.
One takes the AdS metric in the form
$$
ds^2=R^2[-(1+r^2)dt^2+\frac{dr^2}{1+r^2}+r^2d\Omega_3^2]
$$
\be
=R^2[-(1+r^2)dt^2+d{\bf x}d{\bf x}-\frac{({\bf x}d{\bf x})^2}{1+r^2}]
\label{H10a}
\ee
where $r^2={\bf x}{\bf x}$.
One obtains the flat spacetime metric at $R\to\infty$
in  the "proper" coordinates $T=Rt,{\bf X}={\bf x}R$. One denotes
${\bf x}=r\hat {\bf x},~\hat {\bf x}\in S^3$ and for momenta
 ${\bf p}=|{\bf p}|\hat {\bf p}$.

Polchinski \cite{Pol} gave the following prescription for the flat
space $S$-matrix
\be
\langle
{\bf p}_1,\dots,{\bf p}_m; out |{\bf k}_1,
\dots,{\bf k}_n;in\rangle=
\lim_{R\to\infty}Z^{-1}\langle\prod_i\alpha_-(R|{\bf p}_i|,\hat {\bf p}_i)
\prod_j \alpha_+(R|{\bf k}_j|,\hat {\bf k}_j)\rangle
\label{H10}
\ee

Here
\begin{equation}
\alpha_\pm(\omega,\hat {\bf p})
=\lim_{r\to\infty}\int_{\Sigma_r}d\sigma^\mu
(F^{(\pm)}_{\omega\hat {\bf p}}\partial_\mu\Phi
-\partial_\mu F^{(\pm)}_{\omega\hat {\bf p}}\Phi)=
\int dtd^3\hat xf^{(\pm)}_{\omega\hat {\bf p}}(t,\hat {\bf x})
{\phi}(t,
\hat {\bf x})\ .
\label{H11}
\end{equation}

Functions $F^{(\pm)}_{\omega\hat {\bf p}}
(t,{\bf x})$ are classical solutions of
the wave equation $\Box F^{(\pm)}_{\omega\hat {\bf p}}=0$ with the following
properties. In the neighborhood of the origin $r=0$

\be
F^{(\pm)}_{\omega\hat {\bf p}}(t,x)
\approx\psi^{(\pm)}_{\omega\hat {\bf p}}(t,{\bf x})
e^{-i\omega(t-\hat {\bf p}\hat{\bf x})}
\ee
where $\psi^{(\pm)}_{\omega\hat {\bf p}}
(t,{\bf x})$ goes to identity as $R\to\infty$.
So in the flat space limit one gets the plane wave solution.
The asymptotic behaviour at the boundary is
\be
\lim_{r\to\infty}F^{(\pm)}_{\omega\hat {\bf p}}(t,{\bf x})=
f^{(\pm)}_{\omega\hat
{\bf p}}(t,\hat {\bf x})=G^{(\pm)}_\omega\left(t\mp{\pi\over2}\,,|
\hat {\bf x}\mp
\hat {\bf p}|\right)e^{-i\omega t}
\ee
where
\be
G^{(\pm)}_{\omega}(\tau,\theta)=-e^{\pm i\pi\omega/2}\left({2\over
\omega}\right)^{3/2}{1\over\sqrt\pi}\,\exp\left\{-{\omega
\over2}\,(\theta^2+\tau^2)\right\}
\ee
The factor $Z$ is
\be
Z=\int dt d^4x\prod_i\psi^\pm_{\omega_i\hat {\bf p}_i}
(t,{\bf x})
\ee
The operator $\phi(t,\hat {\bf x})$ is defined by the  relation
\be
\lim_{r\to\infty}r^4\Phi(t,{\bf x})=\phi (t,\hat{\bf x})
\ee
and it is interpreted as an operator ${\cal O}$
in the gauge theory on $S^3$.

For quantization one needs a complete orthonormal system
of solutions of the wave equation.
It seems functions $\{F_{\omega \hat{\bf p}} \}$
do not form such a system.
Quantization of the free scalar field in AdS has been performed in
\cite{AIS}. In the coordinates $\mbox{sinh}\chi =\tan \rho$
the mode expansion of scalar field is
$$
\Phi (t,\rho ,\Omega)=
\sum_{m,n,l}[\frac{(l+n+2)(l+n+3)}{(n+1)(n+2)C_l}]^{1/2}
\sin ^l\rho \cos ^4\rho P^{(l+1,2)}_n(\cos 2\rho)
$$
$$
[e^{-i\omega _{ln}t}Y_{lm}(\Omega)a_{mnl}
+h.c.]
$$
Creation and annihilation operators $a^*_{mnl}$, $a_{mnl}$
satisfy the standard Bose commutation relations.
Let us notice here that the Hibert space of the large N limit
is not the ordinary Bose Fock space, but rather it is the quantum Boltzmann
space \cite{AVB,Vol}.The inteprolating field is nonlocal and it should
live int the quantum Boltzmannian Hilbert space. It seems that entangled
commutation relations \cite{AAV} are appropriate for describing collective
degrees of freedom.

The on-shell functional integral representation for $S$-matrix
\cite{AFS} reads
$$
S(\Phi_{in,out})=\int e^{iI(\Phi)}{\cal D}\Phi
$$
One integrates over the field
configurations with the prescribed behaviour at infinity.

It will be convenient to think on the $S$--matrix  in the first
quantized interpretation.
Let us  for simplicity consider a process $2\to2$ in the Born
approximation . The answer is
\begin{equation}
\langle p_1p_2|S|p'_1p'_2\rangle=\int\prod^2_{i=1}
\frac{e^{i\omega(\vec p_i)t-i\vec x\vec p_i}}{(2\pi)^2)\sqrt{2\omega(p_i)}}
\,\cdot\prod^2_{j=1}
\frac{e^{-i\omega(p'_j)t+i\tilde y\vec p_j}}{(2\pi)^2\sqrt{2\omega(p_j)}}
\cdot K(x,y)dx dy\label{FQ1}
\end{equation}
Here we interpret the plane wave as the amplitude
of the probability to find a particle with
given momenta and energy at the point $x$. The kernel $K(x,y)$ is the
transition amplitude from the point $x$ to the point $y$. The plane wave
describes the travelling of free particle along the geodesics.
We consider the pre-$S$-matrix
$<...|{\cal S}(R)|...>$ using an intuitive first quantized approach.
We  take ${\cal S}(R)$ in the form which naturally generalizes the
Feynman flat space formula (\ref{FQ1}),

\be
\langle\omega'_1\hat p'_1,\omega'_2\hat p'_2,\dots,\omega'_m
\hat p'_m|{\cal S}(R)|\omega_1\hat p_1,\dots,\omega_n\hat p_n
\rangle=
\label{H20}
\ee

$$
\int \prod _i \sqrt{g(x_i,t_i)}dx_idt_i~K_{as}(\omega'_i\hat
p'_i|x_i,t_i)K_{as}(x_i,t_i|\omega_{j_i}\hat p_{j_i})+$$

$$
\int \prod _{i} \sqrt{g(x_i,t_i)}dx_idt_i
K_{as}(x_i,t_i|\omega_{j_i}\hat p_{j_i})
\prod _{j}\sqrt{g(x_i,t_i)}dx_idt_i~K_{as}(\omega'_i\hat
p'_i|x'_i,t'_i)
K(x'_1,t'_1,...x_n,t_n)$$
The kernel admits the first quantized representation.
In particular, for $2\to 2$ process the formula (\ref{H20})
becomes

\be
\langle\omega'_1\hat p'_1,\omega'_2\hat p'_2,
|{\cal S}(R)|\omega_1\hat p_1,\omega_2\hat p_2\rangle=
 \label{H21}
\ee

$$
\int
\sqrt{g(x_1,t_1)}dx_1dt_1
\sqrt{g(x_2,t_2)}dx_2dt_2
\sqrt{g(x'_1,t'_1)}dx'_1dt'_1
\sqrt{g(x'_2,t'_2)}dx'_2dt'_2
~K_{as}(\omega'_1\hat p'_1|x'_1,t'_1)\cdot$$
$$
 K_{as}(\omega'_2\hat p'_2|x'_2,t'_2)
 K_{as}(x_1,t_1|\omega_{1}\hat p_{1})
 K_{as}(x_2,t_2|\omega_{2}\hat p_{2})[
 \delta (x_1-x'_1)\delta (t_1-t'_1)+...]
+$$
$$
\sqrt{g(x,t)}dxdt
\sqrt{g(x',t')}dx'dt'
~K_{as}(\omega'_1\hat p'_1|x',t')
 K_{as}(\omega'_2\hat p'_2|x',t')
\cdot
$$
$$
 K_{as}(x,t|\omega_{1}\hat p_{1})
 K_{as}(x,t|\omega_{2}\hat p_{2})K(x',t';x,t)
$$
Here $K(x,t;x',t')$ is the transition amplitude from the point $x,t$
to point
$x',t'$ .
The 2-point transition amplitude includes the sum over
all possible  curves with branches,
\be
K(x,t;x',t')=\sum_{topol}\int _{x,t)}^{x',t')}e^{iS}{\cal D}x {\cal D}t
\label{H23}
\ee

In the Born approximation one performs only summation over
smooth curves
\be
K_{Born}(x,t;x',t')=\int _{x,t)}^{x',t')}e^{iS}{\cal D}x {\cal D}t
\label{H22}
\ee

For $K_{as}(\omega'_1\hat p'_1|x'_1,t'_1)$ one expects
the formula
\be
K_{as}(\omega\hat p|x,t)=K_{as}(y_{as}(\omega\hat p),t_{as}(\omega\hat p)|x,t)
\label{H24}
\ee
Intuitively it is clear that these asymptotic kernel
are solutions of wave equations in the AdS space with special boundary
conditions.
From the previous discussions it seems natural
 to take
\be
K_{as}(\omega\hat p|x,t)=
\frac{F^{(\pm)}_{\omega\hat p}(t,x)}{\psi^\pm_{\omega\hat p}(t,x)}
\label{H25}
\ee

Therefore, for pre-${\cal S}$-matrix we obtain
\be
\langle\omega'_1\hat p'_1,\omega'_2\hat p'_2,\dots,\omega'_m
\hat p'_m|{\cal S}(R)|\omega_1\hat p_1,\dots,\omega_n\hat p_n
\rangle _{connected ~part}=
\label{H26}
\ee

$$
\int \prod _{i}
\frac{F^{(\pm)}_{\omega_i\hat p_i}(t_i,x_i)}
{\psi^\pm_{\omega_i\hat p_i}(t_i,x_i)}
\sqrt{g(x_i,t_i)}dx_idt_i
\prod _{j}
\frac{F^{(\pm)}_{\omega '_i\hat p '_i}(t'_i,x'_i)}
{\psi^\pm_{\omega '_i\hat p'_i}(t'_i,x'_i)}\sqrt{g(x'_i,t'_i)}dx'_idt'_i
K(x'_1,t'_1,...x_n,t_n)
$$
In conclusion, in this note we have given a recipe of how to
encode an information on the flat space $S$-matrix
into the holographic data on the boundary of AdS. Although this recipe
is not very transparent it permits to decode the flat space $S$-matrix.
It would be interesting to study a relation between the recipe
and the operator prescription for the flat space holographic $S$-matrix.
We have proposed that the interpolating field lives in the Boltzmannian
Hilbert space.

\section*{Acknowledgments}

I would like to thank I.V.Volovich
for discussions.  This work was supported in part by
INTAS grant 96-0698.

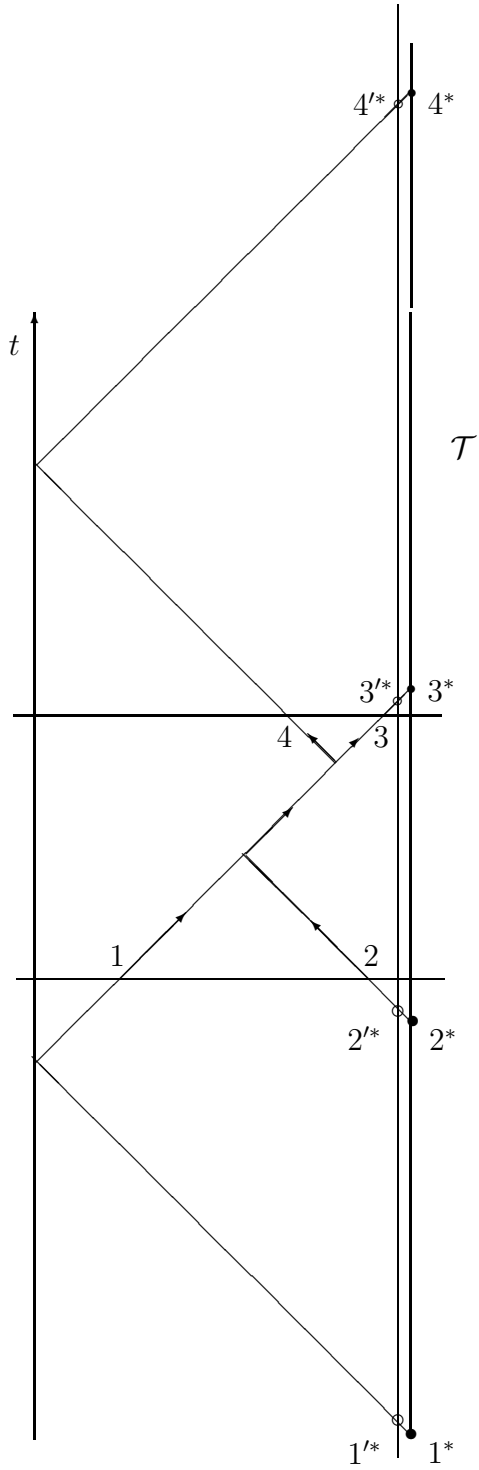
\begin{figure}
\begin{center}
\unitlength=1.00mm
\special{em:linewidth 0.4pt}
\linethickness{0.4pt}
\begin{picture}(62.00,196.00)
\put(5.00,5.00){\vector(0,1){150.00}}
\put(55.00,155.00){\line(0,-1){149.67}}
\put(5.00,55.00){\line(1,1){49.67}}
\put(55.00,5.67){\line(-1,1){50.33}}
\put(54.67,61.00){\line(-1,1){22.00}}
\put(45.00,95.00){\line(-1,1){40.00}}
\put(5.33,134.67){\line(1,1){49.78}}
\put(55.11,155.56){\line(0,1){35.11}}
\put(2.22,101.33){\line(1,0){56.89}}
\put(59.56,66.22){\line(-1,0){56.89}}
\put(16.44,66.44){\vector(1,1){8.67}}
\put(49.33,66.44){\vector(-1,1){7.56}}
\put(41.56,74.22){\line(-1,1){8.44}}
\put(33.11,82.67){\line(0,0){0.00}}
\put(33.11,82.89){\vector(1,1){6.22}}
\put(44.89,95.11){\vector(1,1){3.33}}
\put(45.11,95.11){\vector(-1,1){3.78}}
\put(53.33,196.00){\line(0,-1){193.33}}
\put(53.33,7.56){\circle{1.33}}
\put(55.11,5.78){\circle*{1.33}}
\put(53.33,62.00){\circle{1.33}}
\put(55.33,60.67){\circle*{1.33}}
\put(59.33,58.22){\makebox(0,0)[cc]{$2^*$}}
\put(48.89,58.44){\makebox(0,0)[cc]{$2'^*$}}
\put(48.89,3.33){\makebox(0,0)[cc]{$1'^*$}}
\put(59.11,3.33){\makebox(0,0)[cc]{$1^*$}}
\put(49.78,69.33){\makebox(0,0)[cc]{$2$}}
\put(16.00,69.33){\makebox(0,0)[cc]{$1$}}
\put(38.28,98.56){\makebox(0,0)[cc]{$4$}}
\put(51.11,98.56){\makebox(0,0)[cc]{$3$}}
\put(50.44,104.72){\makebox(0,0)[cc]{$3'^*$}}
\put(59.11,104.72){\makebox(0,0)[cc]{$3^*$}}
\put(53.28,103.22){\circle{1.00}}
\put(55.11,104.89){\circle*{1.00}}
\put(49.63,182.74){\makebox(0,0)[cc]{$4'^*$}}
\put(59.15,182.74){\makebox(0,0)[cc]{$4^*$}}
\put(53.40,182.74){\circle{1.19}}
\put(55.18,184.13){\circle*{1.19}}
\put(2.33,150.67){\makebox(0,0)[cc]{$t$}}
\put(62.00,137.00){\makebox(0,0)[cc]{${\cal T}$}}
\end{picture}

\end{center}
\label{Fig1}
\caption{Born approximation for pre-scattering  in AdS }
\end{figure}

\end{document}